\documentclass[11pt]{article}
\usepackage{graphicx}
\graphicspath{{C:/Users/xinwei/Documents/6EPMapping/gunterPaper/}}

\columnwidth\textwidth

\def\al{\alpha}
\def\ga{\gamma}
\def\ro{\varrho}

\def\d{\partial}
\def\=d{\,{\buildrel\rm def\over =}\,}

\def\sqr#1#2{{\vcenter{\vbox{\hrule height.#2pt\hbox{\vrule width.
#2pt height#1pt \kern#1pt \vrule width.#2pt}\hrule height.#2pt}}}}

\def\e{{\rm em}}
\def\o{{\rm obs}}
\def\te{\vartheta}
\def\B{\Bigl}

\begin{document}

\title{Cosmology with the cosmic rest frame }
\author{G\"unter Scharf
\footnote{e-mail: scharf@physik.uzh.ch}
\\ Physics Institute
\\ University of Z\"urich, 
\\ Winterthurerstr. 190 , CH-8057 Z\"urich, Switzerland}

\date{}

\maketitle\vskip 3cm

\begin{abstract} 

We assume a one-to-one correspondence between comoving coordinates and the cosmic rest frame in a spherically symmetric inhomogeneous universe. This strongly restricts the solutions of Einstein's equations: (i) The pressure must be zero. (ii) The metric does not depend on the radial coordinate, the FLRW cosmology is excluded (strong homogeneity). (iii) The solution corresponds to the homogeneous Datt-Ruban solution. Beside the Hubble constant, it contains one free constant of integration which can be chosen in order to represent the measured Hubble diagram without using a cosmological constant. 


\end{abstract}

\newpage

\section{Introduction}

To understand gravity on the same basis as the other interactions in nature one has to consider it as a spin-2 gauge theory [1]. But then a cosmological constant does not appear in a natural way; Einstein's ``blunder'' remains a blunder. As a consequence we cannot accept the standard FLRW model of cosmology where the mysterious vacuum energy plays a prominent role. In addition the continuing absence of dark matter particles in underground searches increases the doubts in the validity of this model.

In hydrodynamics both comoving Lagrangian and fixed Eulerian coordinates are successfully in use [2]. In cosmology one works almost exclusively with comoving coordinates which follow the motion of the matter. The reason is that Einstein's equations are simpler in comoving coordinates because the Einstein tensor contains much less terms. However the Eulerian coordinates adapted to the CMB rest frame have important merits. First the cosmic rest frame is a global frame for the whole Universe in space and time, the Universe is topologically trivial. The comoving coordinates on the other side only define a patch of local frames. There may arise problems in the large which opens Pandora's box for speculations. Secondly, the cosmic rest frame is observable by measuring the dipole anisotropy of the CMB. As a consequence the interpretation of the solutions of Einstein's equations is much simpler and direct. For example, since the velocity derived from the dipole anisotropy is of the same order of magnitude as the local velocity due to attraction by nearby galaxies, the earth can be considered at rest in the cosmic rest frame. Since the CMB is isotropic to a high degree the earth is even at the center of spherical symmetry in an inhomogeneous but isotropic Universe. This center is taken as the origin $r=0$ of the cosmic rest frame, so that all measured distances agree with the coordinate distance $r$.

In the FLRW model the observable quantities like redshift $z(t)$ and apparent luminosity or magnitude $m(t)$ are functions of the cosmic time $t$ only. Eliminating $t$ one obtains a magnitude - redshift relation $m=m(z)$ from the solution of Einstein's equation and can compare it with measurements. In Einstein's equation the sources of the cosmic gravitational field (matter and radiation) must be put in. In practice one represents the sources by means of free parameters and determines their values by fitting the observations. In this way the hypothetical sources dark matter and dark energy have been introduced.

In inhomogeneous cosmology redshift $z(t,r)$ and magnitude $m(t,r)$ depend on a radial coordinate in addition, which gives the place of the radiating source at the time $t$ of emission. Then it is impossible to eliminate both $t$ and $r$ if only $z(t,r)$ and $m(t,r)$ are known. Without any calculation we see that by assuming an inhomogeneous cosmic gravitational field any Hubble diagram can be represented exactly. Some measured $z$ and $m$ determine $t$ and $r$ of the radiating source. There exist very many solutions describing inhomogeneous gravitational fields. It is rather hopeless to select the right one by comparing with the very few observations. We need a theoretical criterion in addition. We shall obtain such a criterion by considering the cosmic rest frame and the comoving coordinates simultaneously.

In the Einstein anniversary book [1] inhomogeneous cosmology in the cosmic rest frame is treated by a perturbation theory. Here we develop exact methods.  We start by analyzing the geodesics in the cosmic rest frame. The null geodesics which describe the propagation of light have a remarkable property: the light speed is direction dependend. In particular,  considering an expanding universe a radial light ray moving inwards is slower than an outward light ray. This should not be a surprise because the light speed depends on the gravitational fields. This is clearly
seen by the time delay in gravitational lensing, but under the influence of homogeneous cosmology it is often forgotten. The directional dependence of the light speed can be seen in the Michelson - Morley experiment. But in the analysis of this experiment the motion of the observer with respect to the cosmic rest frame must be taken into account. A null result of the experiment in all directions fixes the radial velocity of the observer, so that this experiment has an important cosmological consequence. The radial velocity $u(t,r)$ so obtained is assumed to be true for all matter in the universe so that there is still a democratic principle in our model. Next we ask whether this radial motion of ordinary matter is geodesic. The answer is yes if a first condition on the metric functions is satisfied. This condition is equivalent to zero pressure in agreement with a general theorem [4]. In this way we are lead to consider dust as the source of the cosmic gravitational field. Then our model agrees with the LTB model of inhomogeneous cosmology, but its simultaneous treatment in the cosmic rest frame and comoving coordinates gives new physics. 

The paper is organized as follows. In the next section we study geodesics of radiation and matter and establish our model of zero pressure dust.  In Sect.3 we set up Einstein's equations in the cosmic rest system and discuss the connection with the comoving description. As a basic principle we assume a one-to-one correspondence between comoving coordinates and the cosmic rest frame. The necessary integrability conditions strongly restrict the possible solutions of Einstein's equations. Transforming the Einstein tensor $G_\mu^\nu$ to comoving coordinates in Sect.4 gives further conditions on the metric functions. In Sect.5 we use the known solution in the comoving coordinates to construct a unique physical solution in the cosmic rest system which corresponds to the homogeneous Datt-Ruban solution [4]. In the last two sections the observable consequences of this model are discussed, in particular the magnitude - redshift relation. The measured Hubble diagram can be nicely reproduced, but a cosmological constant and dark matter of high density are excluded.

\section{Geodesics and the energy-momentum tensor}

We choose global spherical coordinates $t,r,\te,\phi$ in the cosmic rest frame and assume a spherically symmetric inhomogeneous cosmic gravitational field given by the line element
$$ds^2=dt^2+2b(t,r)dt\, dr-a^2(t,r)dr^2-c(t,r)^2[d\te^2+\sin^2\te d\phi^2].\eqno(2.1)$$
For $b=0$ the metric is the comoving Bondi - metric [5]. We assume $b\ne 0$ throughout in order to have the cosmic rest frame. The comoving coordinates will be introduced in section 3 in addition. The components of the metric tensor are
$$g_{00}=1,\quad g_{01}=b(t,r),\quad g_{11}=-a^2(t,r)$$
$$g_{22}=-c^2(t,r),\quad g_{33}=-c^2(t,r)\sin^2\te,\eqno(2.2)$$
and zero otherwise. The components of the inverse metric are equal to
$$g^{00}={a^2\over D},\quad g^{01}={b\over D},\quad g^{11}=-{1\over D}\eqno(2.3)$$
$$g^{22}=-{1\over c^2},\quad g^{33}=-{1\over c^2\sin^2\te}\eqno(2.4)$$
where
$$D=a^2+b^2\eqno(2.5)$$
is the determinant of the $2\times 2$ matrix of the $t$, $r$ components. The non-vanishing Christoffel symbols are given by
$$\Gamma^0_{00}={b\dot b\over D},\quad \Gamma^0_{01}=-{ab\over D}\dot a,\quad
\Gamma^0_{11}={1\over D}(a^3\dot a-aba'+a^2b')\eqno(2.6)$$
$$\Gamma^0_{22}={1\over D}(c\dot ca^2+bcc'),\quad\Gamma^0_{33}={\sin^2\te\over D}(c\dot ca^2+bcc')$$
$$\Gamma^1_{00}=-{\dot b\over D},\quad\Gamma^1_{01}={a\dot a\over D},\quad \Gamma^1_{11}={1\over D}(ab\dot a+aa'+bb')$$
$$\Gamma^1_{22}={1\over D}(bc\dot c-cc'),\quad\Gamma^1_{33}={1\over D}(bc\dot c-cc')\sin^2\te,\eqno(2.7)$$
$$\Gamma^2_{02}={\dot c\over c},\quad\Gamma^2_{12}={c'\over c},\quad\Gamma^2_{33}=-\sin\te\cos\te\eqno(2.8)$$
$$\Gamma^3_{03}={\dot c\over c},\quad\Gamma^3_{13}={c'\over c},\quad\Gamma^3_{23}={\cos\te\over\sin\te}.\eqno(2.9)$$
Here the dot means $\d/\d t$ and the prime $\d/\d r$.

First we investigate the radial null geodesics. From $ds^2=0=dt^2+2b(t,r)dt\, dr-a^2(t,r)dr^2$ we obtain
$${dr\over dt}={1\over a^2}(b\pm\sqrt{D})\equiv c_r.\eqno(2.10)$$
Interpreting this as the velocity of light in the cosmic rest system, we see that inward and outward light rays propagate with different velocities. This should not be a surprise because the light speed depends on the inhomogeneous cosmic gravitational field. However this effect is in principle observable by the Michelson - Morley experiment. A null result of this experiment must be due to the motion of the observer in the cosmic rest frame.
Let $c^\mu=(1,c_r,0,0)$ be the 4-velocity of the light ray with
$$g_{\mu\nu}c^\mu c^\nu =0\eqno(2.11)$$ 
and $u^\mu=(u^0,u^1,0,0)$ the observer's velocity with
$$g_{\mu\nu}u^\mu u^\nu=1,\eqno(2.12)$$
then we decompose $c^\mu$ into components parallel and orthogonal to $u^\mu$ [3]
$$c^\mu=(u_\al c^\al)(u^\mu+e^\mu)\eqno(2.13)$$
where
$$e^\mu u_\mu=0.\eqno(2.14)$$
Here
$$u_\al c^\al=u^0\B[1+{b\over a^2}(b\pm\sqrt{D})\B]\mp u^1\sqrt{D}\eqno(2.15)$$
can be interpreted as the speed of the light rays in the observer's rest frame. We demand that the in- and outward light rays have the same speed in the observer's frame
$$u^0\B[1+{b\over a^2}(b+\sqrt{D})\B]-u^1\sqrt{D}=u^0\B[1+{b\over a^2}(b-\sqrt{D})\B]+u^1\sqrt{D}$$
which yields
$$u^0=u^1{a^2\over b}$$
or
$$u^0={a\over\sqrt{D}},\quad u^1={b\over a\sqrt{D}}\eqno(2.16)$$
for the normalized 4-velocity of the observer. For later use we also note the components with lower indices.
$$u_0={\sqrt{D}\over a},\quad u_1=0,\eqno(2.17)$$
the angular components $u^2$ and $u^3$ vanish. The light speed (2.15) in the observer's rest system is now equal to
$$\bar c=\pm{\sqrt{D}\over a}\eqno(2.18)$$
and its magnitude does not depend on the direction.

In the following we shall assume that all matter in the universe is in radial motion with velocity $u^\mu$ (2.16). To check that this velocity is really universal we consider a non-radial light ray with wave vector
$$k^\mu=({\omega\over\bar c}, 0,k_\te, 0).\eqno(2.19)$$
Normalization implies
$$k_\te =\pm{\omega\over\bar cc(t,r)}.$$
With (2.17) we find for the projection onto the observer's rest frame
$$u_\mu k^\mu=u_0{\omega\over\bar c}={\sqrt{D}\over a}{\omega\over\bar c}=\omega$$
which shows that this light ray has the same speed as the radial ones.

Next we are interested in the geodesics for material test bodies which are solutions of the geodesic equation
$${d^2x^\mu\over d\tau^2}+\Gamma^\mu_{\al\beta}{dx^\al\over d\tau}
{d x^\beta\over d\tau}=0.\eqno(2.20)$$
Again we consider radial geodesics $\te={\rm const.}, \phi={\rm const.}$.  Then the geodesic equations are simply given by 
$${d^2t\over d\tau^2}+\Gamma^0_{00}\B({dt\over d\tau}\B)^2+2\Gamma^0_{01}{dt\over d\tau}{dr\over d\tau}+
\Gamma^0_{11}\B({dr\over d\tau}\B)^2=0\eqno(2.21)$$
$${d^2r\over d\tau^2}+\Gamma^1_{00}\B({dt\over d\tau}\B)^2+2\Gamma^1_{01}{dt\over d\tau}{dr\over d \tau}+
\Gamma^1_{11}\B({dr\over d\tau}\B)^2=0.\eqno(2.22)$$
As a first step we eliminate the affine parameter $\tau$. From
$${dr\over dt}={dr\over d\tau}{d\tau\over dt}$$
we find
$${d^2r\over dt^2}=\B[{d^2r\over d\tau^2}{d\tau\over dt}+{dr\over d\tau}{d\over d\tau}\B({dt\over d\tau}\B)^{-1}\B]
{d\tau\over dt}=$$
$$={d^2r\over d\tau^2}\B({d\tau\over dt}\B)^2-{dr\over d\tau}\B({dt\over d\tau}\B)^{-2}{d^2t\over d\tau^2}{d\tau\over dt}=$$
$$={d^2r\over d\tau^2}\B({d\tau\over dt}\B)^2-\B({dt\over d\tau}\B)^{-2}{d^2t\over d\tau^2}{dr\over dt}.\eqno(2.23)$$
Substituting the geodesic equations (2.21) (2.22) inhere we get
$${d^2r\over dt^2}+\B(2\Gamma^1_{01}-\Gamma^0_{00}\B){dr\over dt}+\B(\Gamma^1_{11}-2\Gamma^0_{01}\B)\B({dr\over dt}\B)^2-\Gamma^0_{11}\B({dr\over dt}\B)^3+\Gamma^1_{00}=0.\eqno(2.24)$$
This is now a first order differential equation for the radial velocity
$$v_r(t)={u^1\over u^0}={b\over a^2}={dr\over dt}.\eqno(2.25)$$

We test whether the radial velocity (2.25) is a solution of this geodesic equation (2.24). Calculating the total derivative
$${d^2r\over dt^2}={\dot b\over a^2}-2{b\dot a\over a^3}+\B({b'\over a^2}-2{ba'\over a^3}\B){b\over a^2}$$
and substituting the Christoffel symbols into (2.24) we see that the equation is satisfied if and only if
$${b'\over b}={a'\over a}.\eqno(2.26)$$
This is a first important condition on the metric functions which we shall assume in the following and call it the ab-condition for short. It implies the relation
$$b(t,r)=a(t,r)f_0(t)\eqno(2.27)$$
where $f_0(t)$is an arbitrary function of the cosmic time $t$.
It follows from a general theorem [4] that if matter moves on geodesics then the pressure must be zero. To verify this we consider the energy-momentum tensor of a perfect pressure-less fluid
$$T_m^{\mu\nu}=\ro u^\mu u^\nu,\eqno(2.28)$$
with $u^\mu$ given by (2.16) and $\ro$ is the energy density in the cosmic rest system. We note that the tensor with lower indices has only one non-vanishing component, namely
$$(T_m)_{00}=\ro{D\over a^2}\eqno(2.29)$$
as a consequence of (2.17). Substituting (2.28) into the equation
$$\d_\mu T_m^{\mu\nu}+\Gamma_{\mu\lambda}^\nu T_m^{\mu\lambda}+\Gamma_{\mu\lambda}^\mu T_m^{\nu\lambda}=0\eqno(2.30)$$
of energy-momentum conservation, then for $\nu=0$ we get the equation
$$\dot\ro{a^2\over D}+\ro'{b\over D}+\ro\B[\dot a{a\over D}+{b'\over D}-{a'\over D}{b\over a}+2{\dot c\over c}{a^2\over D}+2{c'\over c}{b\over D}\B]=0$$
and for $\nu=1$ we have 
$$\dot\ro{a^2\over D}+\ro'{b\over D}+\ro{a^2\over b}\B[{\dot a\over D}{b\over a}+2{b'\over D}{b\over a^2}-2{a'\over D}{b^2\over a^3}+2{\dot c\over c}{b\over D}+2{c'\over c}{b^2\over a^2D}\B]=0.$$
Both equations coincide if and only if the above condition (2.26) is satisfied. This condition is necessary and sufficient for zero pressure and for the geodesic motion of the matter. This means that our model in the cosmic rest frame is physically equivalent with the LTB model. However since the latter has only been discussed in comoving coordinates we shall obtain additional new results.

\section{Einstein's equations in the cosmic rest frame}.

From the Christoffel symbols we calculate the Ricci tensor
$$R_{\mu\nu}=\d_\al\Gamma^\al_{\mu\nu}-\d_\nu\Gamma^\al_{\mu\al}+
\Gamma^\al_{\al\beta}\Gamma^\beta_{\mu\nu}-\Gamma^\al_{\nu\beta}
\Gamma^\beta_{\al\mu}\eqno(3.1)$$
For comparison with results in the literature we give the mixed components:
$$R_0^0=-{\ddot a\over D}a-{\dot a^2\over D^2}b^2+{\dot a\dot b\over D^2}ab+{\dot bb'\over D^2}b-{\dot b'\over D}+{a'\dot b\over D^2}a-2{\dot c' 
\over cD}b-$$
$$-2{\ddot c\over c}{a^2\over D}+2{\dot c\dot b\over cD^2}a^2b-2{c'\dot b\over cD^2}a^2-2{\dot a\dot c\over cD^2}ab^2+2{c'\dot a\over cD^2}ab \eqno(3.2)$$
$$R_1^0=-2{\dot c'\over c}{a^2\over D}-2{c''\over c}{b\over D}+2{\dot ac'\over Dc}a+2{\dot cb'\over cD^2}a^2b-2{\dot ca'\over cD^2}ab^2+$$
$$+2{c'b'\over cD^2}b^2+2{c'a'\over cD^2}ab\eqno(3.3)$$
$$R_0^1=-2{\ddot cb\over cD}+2{\dot c'\over cD}+2{\dot c\dot b\over cD^2}b^2+2{\dot c\dot a\over cD^2}ab
-2{c'\dot b\over cD^2}b-2{c'\dot a\over cD^2}a\eqno(3.4)$$
$$R_1^1=-{\ddot a\over D}a-{\dot a^2\over D^2}b^2+{\dot a\dot b\over D^2}ab+{\dot bb'\over D^2}b-{\dot b'\over D}+{a'\dot b\over D^2}a-$$
$$-2{\dot c'b\over cD}+2{c''\over cD}-2{\dot a\dot c\over cD}a+2{\dot c\over cD^2}(a'ab-b'a^2)-2{c'\over cD^2}(a'a+b'b)\eqno(3.5)$$
$$R_2^2=R_3^3=-{\ddot c\over cD}a^2-{\dot a\dot c\over cD^2}(a^3`+2ab^2)+{\dot b\dot c\over cD^2}a^2b+{\dot a c'\over cD^2}ab-{\dot bc'\over cD^2}a^2+$$
$$+{a'\dot c\over cD^2}ab-{b'\dot c\over cD^2}a^2-{a'c'\over cD^2}a-{b'c'\over cD^2}b+{c'^2\over c^2D}+{c''\over cD}-2{\dot c'\over cD}b
-{\dot c^2\over c^2D}a^2-{\dot cc'\over c^2D}b-{1\over c^2}.\eqno(3.6)$$
All other components vanish. The scalar trace is equal to
$$R=R_\mu^\mu=2\B[-{\ddot a\over D}a-{\dot a^2\over D^2}b^2+{\dot a\dot b\over D^2}ab+{\dot bb'\over D^2}b-{\dot b'\over D}+{a'\dot b\over D^2}a\B]-$$
$$-8{\dot c'\over cD}b-4{\ddot c\over cD}a^2+4{\dot b\dot c\over cD^2}a^2b-4{\dot bc'\over cD^2}a^2-4{\dot a\dot c\over cD^2}(a^3+2ab^2)
+4{\dot ac'\over cD^2}ab+$$
$$+4{c''\over cD}+4{a'\dot c\over cD^2}ab-4{b'\dot c\over cD^2}a^2-4{a'c'\over cD^2}a-4{b'c'\over cD^2}b-2{\dot c^2\over c^2D}a^2+2{c'^2\over c^2D}
-2{\dot cc'\over c^2D}b-{2\over c^2}.\eqno(3.7)$$

Next we compute the mixed Einstein tensor
$$G_\mu^\nu=R_\mu^\nu-{1\over 2}\delta_\mu^\nu R.$$
$$G_0^0=2{\dot c'\over cD}b-2{c''\over cD}+2{\dot a\dot c\over cD}a-2{a'\dot c\over cD^2}ab+2{b'\dot c\over cD^2}a^2+$$
$$+2{a'c'\over cD^2}a+2{b'c'\over cD^2}b+{\dot c^2a^2\over c^2D}-{c'^2\over c^2D}+{\dot cc'\over c^2D}b+{1\over c^2}\eqno(3.8)$$
$$G_1^0=R_1^0,\quad G_0^1=R_0^1$$
$$G_1^1=2{\dot c'\over cD}b+2{\ddot c\over cD}a^2+2{\dot a\dot c\over cD^2}ab^2-2{\dot b\dot c\over cD^2}a^2b+2{\dot bc'\over cD^2}a^2-$$
$$-2{\dot ac'\over cD^2}ab+{\dot c^2\over c^2D}a^2-{c'^2\over c^2D}+{\dot cc'\over c^2D}b+{1\over c^2}\eqno(3.9)$$
$$G_2^2={\ddot aa\over D}+{\ddot c\over cD}a^2+{\dot b'\over D}-{c''\over cD}+2{\dot c'\over cD}b+{\dot a\dot c\over cD^2}(a^3+2ab^2)-$$
$$-{\dot b\dot c\over cD^2}a^2b-{\dot ac'\over cD^2}ab+{\dot bc'\over cD^2}a^2-{a'\dot c\over cD^2}ab+{b'\dot c\over cD^2}a^2+{a'c'\over cD^2}a+$$
$$+{b'c'\over cD^2}b+{\dot a^2b^2\over D^2}-{\dot a\dot b\over D^2}ab-{\dot bb'\over D^2}b-{a'\dot b\over D^2}a=G_3^3.\eqno(3.10)$$

Since Einstein's equation in comoving coordinates are integrable [3] [4], one might think that the problem in the cosmic rest frame should be integrable as well. In spite of many attempts we did not find an integrable form for the Einstein tensor above. This is no harm because we can establish the connection between rest frame and comoving coordinates in the following way. We write the line element (2.1) as a sum of squares in the form
$$ds^2=\B({\sqrt{D}\over a}dt\B)^2-\B(a(t,r)dr-{b\over a}dt\B)^2-c^2(t,r)[d\te^2+\sin^2\te d\phi^2]=$$
$$=d\bar t^2-X^2(\bar t,\bar r)d\bar r^2-Y^2(\bar t,\bar r)[d\te^2+\sin^2\te d\phi^2].\eqno(3.11)$$
We use the notation of Bondi [5] to denote the comoving coordinates $\bar x^\mu=(\bar t,\bar r,\bar \te,\bar\phi)$. We note from (3.11) that the angles $\te$ and $\phi$ are the same in the two systems. The 0- and 1-coordinates are related by
$$d\bar t={\sqrt{D}\over a}dt,\quad d\bar r={1\over X}\B(a\,dr-{b\over a}dt\B).\eqno(3.12)$$
The second relation means that $X$ is an integrating denominator for the differential form in brackets. For two variables such an integrating factor always exists without any necessary condition on $a$ and $b$. From (3.12) we find the derivatives
$${\d\bar t\over \d t}={\sqrt{D}\over a}.\quad {\d\bar t\over \d r}=0,\quad {\d\bar r\over \d t}=-{b\over aX},\quad {\d\bar r\over \d r}={a\over X}\eqno(3.13)$$
and the derivatives for the inverse transformation
$${\d t\over\d\bar t}={a\over\sqrt{D}},\quad {\d t\over\d\bar r}=0,\quad {\d r\over\d\bar t}={b\over a\sqrt{D}},\quad {\d r\over\d\bar r}=
{X\over a}.\eqno(3.14)$$
This enables us to define derivatives with respect to the bar-coordinates
$$D_0={\d\over\d\bar t}={a\over\sqrt{D}}\d_0+{b\over a\sqrt{D}}\d_1,\quad D_1={X\over a}\d_1\eqno(3.15)$$
and the inverse relations
$$\d_0={\d\over\d t}={\sqrt{D}\over a}D,_0-{b\over aX}D_1,\quad \d_1={a\over X}D_1.\eqno(3.16)$$

It follows from (3.13) that the comoving time $\bar t$ is a function of the cosmic time $t$ alone. This is in agreement with (2.27) which yields
$${\sqrt{D}\over a}=\sqrt{1+f_0(t)^2}.$$
If $f_0(t)$ is known we can calculate
$$\bar t=\int\limits^{t}\sqrt{1+f_0^2(t')}dt'.\eqno(3.17)$$

We assume a one-to-one correspondence between the two reference systems which we call our basic assumption for short. At the moment this is rather vague, it will become precise if the consequences are worked out. With the basic assumption it must be possible to calculate $\bar r(t,r)$ as a line integral
$$\bar r(t,r)=\int\B({\d\bar r\over \d r}dr+{\d\bar r\over\d t}dt\B)=\int\B({a\over X}dr-{b\over aX}dt\B)\eqno(3.18)$$
and also
$$ r(\bar t,\bar r)=\int\B({\d r\over \d\bar r}d\bar r+{\d r\over\d\bar t}d\bar t\B)=\int\B({X\over a}d\bar r+{b\over a\sqrt{D}}d\bar t\B).\eqno(3.19)$$
This requires the following integrability conditions
$${\rm curl\, grad}\,\bar r(t,r)=0,\quad {\rm curl\, grad}\, r(\bar t,\bar r)=0.$$
These two conditions imply the ab-relation, i.e. zero pressure, and the further condition
$${D_0X\over X}={\d\over\d\bar t}\log X={\dot a\over\sqrt{D}}.\eqno(3.20)$$
We write the right side as
$${\dot a\over a}{a\over\sqrt{D}}={\dot a\over a}{\d t\over\d\bar t}={\d\over\d\bar t}\log a={\d\over\d\bar t}\log X,$$
yielding
$$a=X(\bar t,\bar r)f_1(\bar r)\eqno(3.21)$$
where $f_1$ is some function of $\bar r$ alone.

Our basic assumption has further consequences. The partial derivatives (3.15), (3.16) must commute
$$\d_1\d_0=\d_0\d_1,\quad D_1D_0=D_0D_1.\eqno(3.22)$$
From the first condition we obtain
$$D_1D_0-D_0D_1={b\over\sqrt{D}}{D_1X\over X^2}D_1.\eqno(3.23)$$
So both conditions (3.22) are satisfied if either $b=0$ or
$$D_1X=0.\eqno(3.24)$$
Since we require $b\ne 0$ throughout in order to have the cosmic rest frame, it follows that $X(\bar t)$ depends on time only. This is one half of homogeneity of the universe.

\section{Transformation to comoving coordinates}

To see that the bar-coordinates are really the comoving coordinates we transform the radial 4-velocity (2.16)
$$\bar u^\mu =u^\nu {\d \bar x^\mu\over \d x^\nu}=(1,0,0,0)\eqno(4.1)$$
which is indeed the definition of the comoving coordinates. The transformation of the Einstein tensor
$$\bar G_\mu^\nu (\bar t,\bar r,\te,\phi)=G_\beta^\al(t,r,\te,\phi){\d\bar x^\nu\over\d x^\al}{\d x^\beta\over\d\bar x^\mu}\eqno(4.2)$$
looks equally simple, but the right-hand side is full of rest system quantities with unknown dependence on comoving coordinates. One exception is
$$Y(\bar t,\bar r)=c(t,r)\eqno(4.3)$$
which follows from (3.11) because the angles are not transformed. For the definition of the angles we refer to the discussion given by Bondi [5]. We shall write the derivatives $D_0$, $D_1$ with respect to comoving coordinates with lower indices and comma, dot and prime remain the derivatives in the cosmic rest system, so that the constraint (3.20) is written as
$${\dot a\over\sqrt{D}}={X,_0\over X}.\eqno(4.4)$$

The transformation is most simple for the component
$$G_1^0=R_1^0={2\over cD}(-\dot c'a^2-c''b+\dot ac'a+a'c'{b\over a})\eqno(4.5)$$
where we have simplified (3.3) by means of the ab-relation. Using (3.16) we have for the mixed derivative
$$\d_0\d_1={\sqrt{D}\over X}D_0D_1-{b\over X^2}D_1^2+{bX,_1\over X^3}D_1\eqno(4.6)$$
and
$$\d_1^2={a^2\over X^2}D_1^2+{a\over X}\B({a,_1\over X}-{aX,_1\over X^2}\B)D_1.\eqno(4.7)$$
This gives
$$G_1^0={2\over YD}\B[-a^2{\sqrt{D}\over X}Y,_{01}+{aa,_0\over X}\sqrt{D}Y,_1-{ab\over X^2}a,_1Y,_1\B]$$
Now we obtain
$$\bar G_1^0=G_1^0{\sqrt{D}\over a^2}X={2\over Y}\B(-Y,_{01}-{b\over a}{a,_1\over X\sqrt{D}}Y,_1+{a,_0\over a},Y_1\B)=\eqno(4.8)$$
$$={2\over Y}\B(-Y,_{01}+{X,_0\over X}Y,_1\B)\eqno(4.9)$$
where (4.4) and (3.16) have been used. This is precisely the component $\bar G_1^0$ in the comoving system (see (4.12).
For the convenience of the reader we list the standard comoving components as given by Bondi [5] (his equation (5)):
$$\bar G_0^0=2{X,_0Y,_0\over XY}+{1+Y,_0^2\over Y^2}-{1\over X^2}\B(2{Y,_{11}\over Y}+{Y,_1^2\over Y^2}-2{X,_1Y,_1\over XY}\B)\eqno(4.10)$$
$$\bar G_1^1=2{Y,_{00}\over Y}+{1+Y,_0^2\over Y^2}-{Y,_1^2\over X^2Y^2}\eqno(4.11)$$
$$\bar G_2^2={X,_{00}\over X}+{Y,_{00}\over Y}+{X,_0Y,_0\over XY}-{1\over X^2}\B({Y,_{11}\over Y}-{X,_1Y,_1\over XY}\B)=\bar G_3^3\eqno(4.12)$$
$$\bar G_1^0=-2\B({Y,_{01}\over Y}-{X,_0Y,_1\over XY}\B).\eqno(4.13)$$

Next we transform $G_2^2$ where we need second time derivatives. In (4.4)
$$\dot a=\sqrt{D}{X,_0\over X}$$
we differentiate $\d_t$ which on $X$ is applied in the form (3.16):
$$\ddot a={\dot bb\over\sqrt{D}}{X,_0\over x}+{X,_0^2\over X^2}a+{1\over X}\B({D\over a}X,_{00}-{b\sqrt{D}\over aX}D_1D_0X\B)-$$
$$-{D\over aX^2}X,_0^2+{b\sqrt{D}\over aX^3}X,_0X,_1.\eqno(4.14)$$
In addition we use
$$\dot a'={a\over X}\B[{aa,_1+bb,_1\over\sqrt{D}}{X,_0\over X}+\B({D_1D_0X\over X}-{X,_0X,_1\over X^2}\B)\sqrt{D}\B]\eqno(4.15)$$
and
$$\dot b'=\d_t\B({a'\over a}b)={\dot a'\over a}b+{a'\over a}\dot b-{a'\over a^2}\dot ab=$$
$$={b\sqrt{D}\over X^2}\B(D_1D_0X-{X,_0X,_1\over X}\B)+{\dot b\over a}{a\over X}a,_1.\eqno(4.16)$$
Then we finally arrive at
$$G_2^2={X,_{00}\over X}+{Y,_{00}\over Y}+{X,_0Y,_0\over XY}-{1\over X^2}\B({Y,_{11}\over Y}-{X,_1Y,_1\over XY}\B)-{X,_1Y,_1\over X^3Y}{b^2\over D}.$$
Apart from the last term this is the comoving tensor component $\bar G_2^2$ (4.12). Since the last term vanishes in virtue of (3.24) this component also transforms correctly.

The component $G_1^1$ (3.9) we transform in the same way with the following final result
$$G_1^1={2\over Y}D_0^2Y-2{b\over XY\sqrt{D}}D_1D_0Y+2b{X,_0Y,_1\over X^2Y\sqrt{D}}-2{b^3\over aD^{3/2}}{a,_1Y,_0\over XY}+$$
$$+2{b^2\over D^{3/2}}{b,_1Y,_0\over XY}+{Y,_0^2\over Y^2}-b{Y,_0Y,_1\over XY^2\sqrt{D}}-{a^2Y_,1^2\over X^2Y^2D}+{1\over Y^2}.\eqno(4.17)$$
Then the corresponding comoving quantity becomes
$$\tilde G_1^1=G_1¹-{b\over a^2}G_1^0=2{Y,_{00}\over Y}+{1+Y,_0^2\over Y^2}-{Y,_1^2\over X^2Y^2}+$$
$$+{b²Y,_1^2\over X^2Y^2D}-b{Y,_0Y,_1\over XY^2\sqrt{D}}\eqno(4.18)$$
where the ab-relation was used again. The first line is in agreement with (4.11), so the second line must vanish by our basic principle.
The remaining two terms vanish if either
$${b\over\sqrt{D}}=X{Y,_0\over Y,_1}\eqno(4.19)$$
or
$$Y,_1=0.\eqno(4.20)$$
The second condition implies strong homogeneity, $Y=Y(\bar t)$ is a function of time alone as $X(\bar t)$ (3.24). Note that in the standard FLRW cosmology $Y=\bar r a(\bar t)$ is proportional to the radial coordinate. Since this is usually called homogeneous we call (4.20) strongly homogeneous for distinction.
 
The first condition (4.19) leads to a contradiction as can be seen as follows. It follows from (2.27) that the left side is a function of $t$ only, hence the right side depends on $\bar t$ only, the radial coordinate $\bar r$ must drop out. On the other hand the known solution of the LTB model [4] (p.298) shows that this is only possible if $Y$is of the form
$$Y(\bar t,\bar r)=q(\bar r)\bar t^{2/3}\eqno(4.21)$$
where
$$q={\rm const.}e^{\beta\bar r}.\eqno(4.22)$$
In this case we have 
$$X(\bar t)=Y,_1=q'\bar t^{2/3}$$
which gives $q'=$ const. which contradicts (4.22).

The same two conditions (4.19-20) are found in the transformation of $G_0^0$ which gives the matter density $8\pi G\ro$ ($G$ is Newton's constant). Since $\ro$ is a scalar it must be the same in the two systems. Consequently a non-trivial solution in both systems is only possible with the strongly homogeneous metric functions $X(\bar t)$, $Y(\bar t)$.

\section{Integration of Einstein's equations}

It is our aim to solve Einstein's equation
$$G_\mu^\nu=8\pi GT_\mu^\nu\eqno(5.1)$$
in the cosmic rest frame with the energy-momentum tensor (2.28) for pressure-less dust. Then from the components (3.8-10) we have
$$G_0^0=8\pi G\ro\eqno(5.2)$$
and all other components must be set equal to zero. The integration process consists of two steps: The first is the integration of bar-equations (4.13-15) in comoving coordinates which can be found in the literature as the LTB model [4] [5]. The second step is the calculation of the quantities in the cosmic rest frame and the test of all compatibility conditions.

In the case $Y,_1=0$ Einstein's equations have already been solved in 1938 by B.Datt [6]. This solution is discussed in Ref.[4] and called Datt-Ruban solution. We simply say Datt solution because the later geometric discussion by Ruban is not relevant for our purposes. $\bar G_1^0$ is identically zero in the homogeneous case.  Since $Y(\bar t)$ depends only on $\bar t$ we have
$$\bar G_1^1=2{Y,_{00}\over Y}+{Y,_0^2\over Y^2}+{1\over Y^2}=0.\eqno(5.3)$$
We multiply by $Y^2Y,_0$
$$2Y,_{00}Y,_0Y+Y,_0^3+Y,_0=D_0(Y,_0^2Y+Y)=0\eqno(5.4)$$
and conclude that the bracket is constant $=T_L$, say. This gives
$$Y,_0^2={T_L\over Y}-1.\eqno(5.5)$$
We shall shortly see that $T_L$ determines the lifetime of our re-collapsing universe. This equation can be integrated in the form
$$\bar t-\bar t_0=\int\limits_0^Y {dy\over\sqrt{T_L/y-1}}=\int\limits_0^Y{y\,dy\over \sqrt{T_Ly-y^2}}.\eqno(5.6)$$
Choosing the integration constant $\bar t_0=0$ fixes the origin of the comoving time at $\bar t=0$. The integral (5.6) is most easily solved by the substitution
$$y=T_L\sin^2w.\eqno(5.7)$$
which yields
$$\bar t=T_L(w-\sin w\cos w),\eqno(5.8)$$
where
$$\sin w=\sqrt{Y\over T_L}\eqno(5.9)$$
as a consequence of (5.7). These two equations give $Y(\bar t)$ in parametric form.

Next we consider 
$$\bar G_2^2={X,_{00}\over X}+{Y,_{00}\over Y}+{X,_0Y,_0\over XY}=0.\eqno(5.10)$$
Since $\bar t$ does not explicitly appear we use $Y$ as the independent variable instead:
$$X,_0={\d X\over\d\bar t}={\d X\over\d Y}Y,_0\eqno(5.11)$$
$$X,_{00}={\d^2X\over\d Y^2}\B({T_L\over Y}-1\B)-{\d X\over\d Y}\B({Y,_0^2\over 2Y}+{1\over 2Y}\B)\eqno(5.12)$$
where (5.9) has been used. After multiplication of (5.10) by $XY^3$ we arrive at
$$Y^2(T_L-Y){\d^2 X\over\d Y^2}+({T_L\over 2}Y-Y^2){\d X\over\d Y}-{T_L\over 2}X=0.\eqno(5.13)$$
This equation is again greatly simplified by the above substitution (5.9)
$$Y=T_L\sin^2w.\eqno(5.18)$$
We use
$${\d X\over\d Y}={\d X\over\d w}\B({\d Y\over\d w}\B)^{-1}={1\over 2T_L\sin w\cos w}{\d X\over \d w}\eqno(5.15)$$
and
$${\d^2 X\over\d Y^2}={1\over 4T_L^2\sin^2w\cos^2w}{\d^2X\over\d w^2}-{\cos^2w-\sin^2w\over 4T_L^2\sin^3w\cos^3w}{\d X\over\d w}\eqno(5.16)$$
and finally obtain the simple equation
$$\sin^2w{\d^2 X\over\d w^2}=2X(\bar t).\eqno(5.17)$$
This linear ordinary differential equation was given by Datt [6]. One obvious solution is $\cot w$. Then the second fundamental solution can be obtained by a product ansatz $X=U(w)\cot w$, so that the general solution becomes
$$X(\bar t)=Q\cot w+P(1-w\cot w),\eqno(5.18)$$
where $P$ and $Q$ are constants of integration. In (5.18) we have used the same notation as Datt, except our use of $w$ for the fundamental parameter instead of Datt's $z$, because $z$ is reserved for the redshift in the next section. In the inhomogeneous situation considered by Datt $P$ and $Q$ are arbitrary functions of $\bar r$ which are not restricted by Einstein's equations.

We now turn to the solution of Einstein's equations in the cosmic rest frame which will give more explicit and detailed results. Again we start with $G_0^1=0$ which in the homogeneous situation gives the simple equation
$$-\ddot cbD+\dot c\dot bb^2+\dot c\dot aab=0.\eqno(5.19)$$
In the form
$${\ddot c\over\dot c}={\dot aa+\dot bb\over D}={1\over 2}{d\over dt}\log D={d\over dt}\log\dot c\eqno(5.20)$$
it is integrable and gives
$$\dot c=\ga\sqrt{D}\eqno(5.21)$$
where $\ga$ is a constant of integration.

To obtain $c(t)$ we substitute the result (5.21) into $G_1^1=0$. After multiplying with $c^2$ we find
$$c{2\ga\over D^{3/2}}\dot aa^2(a+b)=-\ga^2a^2-1$$
yielding the desired
$$c=-{a^2\ga^2+1\over\dot aa^2(a+b)}{D^{3/2}\over 2\ga}.\eqno(5.22)$$
On the other hand using the comoving $Y$ we have by (5.21)
$$\dot c={\sqrt{D}\over a}Y,_0=\ga\sqrt{D}\eqno(5.23)$$
which implies
$$Y,_0=\ga a.\eqno(5.24)$$
This shows that $a(t)$ does not depend on $r$, too, so the $f_1$ in (3.21) is a constant:
$$Y,_0=\ga f_1X.\eqno(5.25)$$

Using the comoving solution (5.5) we still obtain more explicit results:
$$Y,_0^2=\ga^2a^2={T_L\over Y}-1={T_L\over c}-1,$$
hence
$$c(t)={T_T\over\ga^2a^2+1}.\eqno(5.26)$$
Substitution into (5.22) leads to the relations
$$T_L=-{(\ga^2a^2+1)^2\over\dot aa^2(a+b)}{D^{3/2}\over 2\ga}\eqno(5.27)$$
$$\dot a=-{\ga^2a^2+1)^2\over a^2(a+b)}{D^{3/2}\over 2T_L\ga}.\eqno(5.28)$$
Differentiating (5.26) and equating it with (5.21)leads to
$${\dot a\over (\ga^2a²+1)^2}=-{\sqrt{D}\over 2T_L\ga a}.$$
Comparing this with (5.28) gives a relation between $a$ and $b$ only with the simple result
$$b(t)=a(t).\eqno(5.29)$$
As a consequence
$$f_0(t)=1$$
in (2.27) and the comoving time (3.17) is simply proportional to the cosmic time
$$\bar t=\sqrt{2}t.\eqno(5.30)$$

Now (5.28) can be completely integrated
$${\dot a\over (\ga^2a²+1)^2}=-{1\over\sqrt{2}T_L\ga}\eqno(5.31)$$
yielding
$${\ga a\over 2(\ga^2a^2+1)}+{1\over 2}\arctan(\ga a)=-{t\over\sqrt{2}T_L}.\eqno(5.32)$$
Here we have chosen the origin of cosmic time at $t=0$ where $a=0$ which is the Big Bang as in the standard cosmology. Note that $\ga<0$. One easily sees from (5.28) that $a(t)$ is monotonely increasing, but it becomes infinite after a finite time $t_L$
$$t_L=\sqrt{2}{\pi\over 4}T_L.\eqno(5.33)$$
This is not a big crunch but an infinite expansion $a(t_L)=\infty$, $L$ stands for lifetime..

With these results one easily checks that the equation $G_2^2=0$ is satisfied. The equation for $G_0^0$ is discussed at the end of the last section.

\section{Magnitude - redshift relation }

We perform the further calculations in the cosmic rest frame because we have more explicit results here. For the redshift we need the radial null geodesics. They are obtained from the geodesic equation
$$k^\mu(\d_\mu k^\nu+\Gamma_{\mu\lambda}^\nu k^\lambda)=0.\eqno(6.1)$$
For $\nu=0$ we have
$$k^0\d_0k^0+\Gamma_{00}^0(k^0)^2+2\Gamma_{01}^0k^0k^1+\Gamma_{11}^0(k^1)^2=0.\eqno(6.2)$$
Here we insert
$$k^1=k^0{dr\over dt}=k^0{1\over a^2}(b-\sqrt{D})\eqno(6.3)$$
where we have chosen a minus sign of the square root in order to have an incoming light ray which propagates from big $r$ towards $r=0$.
The resulting equation can be written as
$${\dot k^0\over k^0}=-{\dot aa+\dot bb\over D}=-{1\over 2}{d\over dt}\log D$$
yielding
$$k^0={C_0\over\sqrt{D}}\eqno(6.4)$$
where $C_0$ is a constant of integration. The spatial component follows from (6.3)
$$k^1={C_0\over a^2}\B({b\over\sqrt{D}}-1\B).\eqno(6.5)$$

For the redshift we must compute
$$u_\mu k^\mu=u_0k^0={\sqrt{D}\over a}{1\over\sqrt{D}}={1\over a}\eqno(6.6)$$
where $u_\mu$is the 4-velocity (2.17) of the emitter ($\e$) and observer ($\o$). Now the redshift is given by
$$1+z={(u_\mu k^\mu)_\e\over (u_\mu k^\mu)_\o}={a(t_\o)\over a(t_\e)}\eqno(6.7)$$
in agreement with standard cosmology. To get contact with observations we substitute
$$da=-{a_\o\over (1+z)^2}dz\eqno(6.8)$$
into (5.31)
$$dt=-{\sqrt{2}T_L\over\ga^2(a^2+1/\ga^2)^2}\,da=$$
$$={\sqrt{2}T_La_\o\over\ga^3}{(1+z)^2\over (a_\o^2+(1+z)^2/\ga^2)^2}\,dz.\eqno(6.9)$$
This yields
$${dz\over dt}={\ga^3\over\sqrt{2}T_La_\o}{(a_\o^2+(1+z)^2/\ga^2)^2\over (1+z)^2}.\eqno(6.10)$$
For $z=0$ this gives the Hubble constant
$${dz\over dt}\B\vert_{z=0}={\ga^3\over\sqrt{2}T_La_\o}(a_\o^2+1/\ga^2)^2=-H_0.\eqno(6.11)$$
Note that $\ga<0$. The second derivative
$${d^2z\over dt^2}={dz\over dt}{\ga^3\over \sqrt{2}T_La_\o}\B[-{2\over (1+z)^3}\B(a_\o^2+{(1+z)^2\over\ga^2}\B)^2+$$
$$+{2\over (1+z)^2}\B(a_\o^2+{(1+z)^2\over\ga^2}\B){2\over\ga^2}(1+z)\B]$$
at $z=0$ gives the deceleration parameter $q_0$
$${d^2z\over dt^2}\B\vert_{z=0}=H_0^2(2+q_0)=$$
$$=H_0^2\B(2-{4\over\ga^2a_\o^2+1}\B)\eqno(6.12)$$
where we have used the expression (6.11) for the Hubble constant. This implies
$$q_0=-{4\over\ga^2a_\o^2+1}=-{4\al^2\over\al^2+1}.\eqno(6.13)$$
The parameter
$$\al^2={1\over\ga^2a_\o^2}\eqno(6.14)$$
plays a fundamental role in the following. Since $q_0$ is negative our theory naturally explains the present acceleration of the expansion without assuming some dark energy.

Finally we want to calculate the luminosity distance $d_L(z)$ and the magnitude $m(z)$. We start from the radial null geodesic in the form
$${dr\over dt}={1\over a^2}(\sqrt{D}-b)={\sqrt{2}-1\over a}=c_0\eqno(6.15)$$
where $c_0$ denotes the local light speed. By integration from the time of emission $t(z)$ to the time $t_0=t_\o$ of observation we get the radial distance
$$R(z)=\int\limits_{t(z)}^{t_0}{\sqrt{2}-1\over a(t)}dt=-(\sqrt{2}-1){\sqrt{2}T_L\over\ga^3}\int{da\over a(a^2+1/\ga^2)^2}=$$
$$=(\sqrt{2}-1){\sqrt{2}T_L\over\ga^3}\int\limits_z^0{dz\over (1+z)(a_\o^2/(1+z)^2+1/\ga^2)^2}$$
where (6.8) was used. With the new variable of integration $x=1/(1+z)$ we finally obtain
$$R(z)={c_0\over H_0}(1+\al^2)^2\int\limits_{1/(1+z)}^1{dx\over x(x^2+\al^2)^2}.\eqno(6.16)$$
Here the present light speed (6.15) and the Hubble constant (6.11) has been inserted and the dimensionless parameter $\al$ (6.14) appears again. The luminosity distance is equal to $(1+z)R(z)$ apart from a constant factor $a_\o$ which is irrelevant for the magnitude below [7][8]. The rational integral in (6.16) is elementary so that
$$d_L(z)={c_0\over H_0}(1+z){(1+\al^2)^2\over 2\al^2}\B[{1\over 1+\al^2}-{(1+z)^2\over 1+\al^2(1+z)^2}+$$
$$+{1\over\al^2}\log{1+\al^2(1+z)^2\over 1+\al^2}\B].\eqno(6.17)$$
The magnitude $m(z)$ is defined by
$$m(z)=5\log_{10}d_L+M+25\eqno(6.18)$$
where $M$ is the absolute magnitude of the supernova standard candle. In the Hubble diagram one plots the distance modulus
$$\mu(z)=m(z)-M.\eqno(6.19)$$
In the next section we shall use the measured value of $\mu$ at $z=1$ to determine the unknown parameter $\al^2$ in (6.17). Then we obtain an excellent representation of the entire Hubble diagram.

We want also calculate the look-back time [8]
$$t(z)=t_\o-t_\e={\sqrt{2}T_La_\o\over\vert\ga^3\vert}\int\limits_0^z{(1+z')^2\over (a_\o^2+(1+z')^2/\ga^2)^2}\,dz'=$$
$$={(1+\al^2)^2\over H_0}\int\limits_{1/(1+z)}^1{dx\over (x^2+\al^2)^2}=$$
$$={1\over H_0}{(1+\al^2)^2\over 2\al^2}\B[{1\over 1+\al^2}-{1+z\over 1+\al^2(1+z)^2}+$$
$$+{1\over\al}\B(\arctan{1\over\al}-\arctan{1\over\al(1+z)}\B)\B].\eqno(6.20)$$
Since we know $\al$ we can determine the temporal structure of the Universe in the next section.

\section{Discussion}

The measured Hubble diagram is nicely represented by the standard FLRW luminosity distance
$$\tilde d_L(z)={c_0(1+z)\over H_0}\int\limits_1^{1+z}{dx\over\sqrt{\Omega_Mx^3+\Omega_\Lambda}}.\eqno(7.1)$$
From the type Ia supernovae observations one has obtained the following parameter values $\Omega_M=0.27$, $\Omega_\Lambda=0.73$, $H_0=72$ km/(s Mpc). This is the best fit in [9]. In the table the corresponding distance modulus $\tilde\mu(z)$ is listed in the second column. We have taken the value
$\tilde\mu(1)=44.08$ at $z=1$ as a measured value and have determined the free parameter $\al^2$ in (6.17-19) such that this value is reproduced. The result is
$$\al^2=6.71.\eqno(7.2)$$
With this value the entire Hubble diagram until $z=10$ is excellently represented by (6.17-19) as can be seen in the third column of the table. The deceleration parameter (6.13) comes out to be
$$q_0=-3.48.\eqno(7.3)$$

$$\vbox{\halign{#&#&#&#&#&#&#&#\cr
$z$\qquad\quad&$\tilde\mu(z)({\rm mag})\qquad$&$\mu(z)({\rm mag})\qquad$&$\tilde t(z)(10^9Y)\qquad\qquad$&$t(z)(10^9Y)\qquad$\cr
\noalign{\hrule\vskip 0.2 cm}
0.01&33.12&33.12&0.1349&0.1348\cr
0.02&34.64&34.64&0.2678&0.2676\cr
0.03&35.53&35.54&0.3990&0.3985\cr
0.04&36.17&36.18&0.5283&0.5275\cr
0.05&36.67&36.68&0.6558&0.6546\cr
0.06&37.08&37.09&0.7816&0.7799\cr
0.07&37.43&37.44&0.9057&0.9034\cr
0.08&37.74&37.75&1.0281&1.025\cr
0.09&38.01&38.02&1.1488&1.145\cr
0.1&38.25&38.26&1.2679&1.263\cr
0.2&39.89&39.91&2.3756&2.360\cr
0.3&40.89&40.91&3.3443&3.317\cr
0.4&41.62&41.64&4.1969&4.158\cr
0.5&42.20&42.22&4.9489&4.903\cr
0.6&42.69&42.71&5.6145&5.565\cr
0.7&43.10&43.12&6.2054&6.157\cr
0.8&43.46&43.48&6.7317&6.690\cr
0.9&43.79&43.80&7.2020&7.171\cr
1.0&44.08&44.08&7.02356&7.608\cr
2.0&46.05&45.96&10.181&10.45\cr
3.0&47.22&47.03&11.318&11.90\cr
4.0&48.05&47.78&11.928&12.79\cr
5.0&48.70&48.36&12.300&13.38\cr
6.0&49.22&48.82&12.541&13.80\cr
7.0&49.67&49.21&12.711&14.12\cr
8.0&50.05&49.55&12.836&14.37\cr
9.0&50.38&49.84&12.93&14.57\cr
10.0&50.68&50.10&13.033&14.73\cr
}}$$

Regarding the magnitude-redshift relation our results show that the Hubble diagram is not very specific for a particular cosmological model. The situation is slightly better for the look-back time $t(z)$ (6.20). In the forth column we list the results for the standard FLRW cosmology [8]
$$\tilde t(z)={1\over H_0}\int\limits_1^{1+z}{dx\over x\sqrt{\Omega_Mx^3+\Omega_\Lambda}}.\eqno(7.4)$$. In the last column the corresponding values of $t(z)$ (6.20) are given. We find increasing differences for $z>1$. Unfortunately the look-back time is not directly measurable.

A time of particular interest is the life-time $t_L$ (5.33). From (6.11) we get
$$T_L=\vert\ga^3\vert{(a_\o^2+1/\ga^2)^2\over\sqrt{2}a_\o H_0}={(1+\al^2)^2\over\sqrt{2}\al^3}T_H=2.42\,T_H.\eqno(7.5)$$
Then the life-time comes out to be
$$t_L=\sqrt{2}{\pi\over 4}T_L=2.69\,T_H\eqno(7.6)$$
which means that our Universe has still $1.69\,T_H$ time to expand.

Before ending an essential point must be discussed. We now consider the last Einstein's equation (5.2) for $G_0^0$. By strong homogeneity there is a great simplification
$$G_0^0=2{\dot a\dot c\over cD}+{\dot c^2a^2\over c^2D}+{1\over c}.\eqno(7.7)$$
From (5.21-22) we have
$${\dot c\over c}=-{2\ga^2\dot aa\over\ga^2a+1}\eqno(7.8)$$
and $\dot a$ is given by (5.31).After substituting this into (7.7) we find
$$G_0^0=0\eqno(7.9)$$
that means the matter density $\ro=0$. This should not be a surprise because we have assumed in Sect.2 that the matter moves on radial geodesics. {\it Then the matter is considered as test particles}, and consequently, the matter density is so small that it does not act as a source of the gravitational field. Summing up our basic principle has finally led to the following non-standard cosmology: The expansion of the Universe on the largest scale must be described by a vacuum solution of Einstein's equations. Matter should be considered as an inhomogeneous perturbation of this background. Dark matter of high density is excluded. The failure of Newtonian gravity in the rotation curves of galaxies is discussed in [1].

\end{document}